\documentclass[pre, amssymb, amsmath, nofootinbib, 10pt, twocolumn]{revtex4}
	
	\usepackage{graphicx}
	\usepackage{color}
	\usepackage{verbatim}
	\usepackage{subfigure}
	\usepackage{amssymb}
\usepackage[normalem]{ulem}  

\begin{document}

	\title{Non-equilibrium work energy relation for non-Hamiltonian dynamics}
	\author{Dibyendu Mandal$^1$ and Michael R. DeWeese$^{1,2}$}
	\affiliation{$^1$Department of Physics, University of California, Berkeley, CA 94720 \\
	$^2$Redwood Center for Theoretical Neuroscience and Helen Wills Neuroscience Institute, University of California, Berkeley CA 94720}

	\begin{abstract}

	Recent years have witnessed major advances in our understanding of nonequilibrium processes. 
	The Jarzynski equality, for example, provides a link between equilibrium free energy differences and finite-time, nonequilibrium dynamics. 
	We propose a generalization of this relation to non-Hamiltonian dynamics, relevant for active matter systems, continuous feedback, and computer simulation. 
	Surprisingly, this relation allows us to calculate the free energy difference between the desired initial and final equilibrium states using arbitrary dynamics. 
	As a practical matter, this dissociation between the dynamics and the initial and final states promises to facilitate a range of techniques for free energy estimation in a single, universal expression.

	\end{abstract}

	\maketitle

	\section{Introduction} 
	
	Free energy determination lies at the heart of nearly any application of statistical mechanics~\cite{Chandler1987, Chaikin1995}, the conventional methods being based on either the calculation of a partition function or the determination of work in a transition from one equilibrium state to another~\cite{Chipot2007}.  
	In the latter case, the Helmholtz free energy $F_0(A, \beta)$ of the initial equilibrium state with probability density $\rho_A^\text{eq} \propto \exp{- \beta{\cal H}_A}$ is assumed to be known for a given Hamiltonian ${\cal H}_A$ with external parameters $A$ and inverse temperature $\beta = 1 / k_\text{B} T$, where $k_\text{B}$ denotes the Boltzmann constant. 
	Then, the free energy difference $\Delta F_0 \equiv \left[F_0 (B, \beta) - F_0(A, \beta) \right]$ corresponding to the transition to another Hamiltonian ${\cal H}_B$ is estimated from the work as the external parameters are switched from $A$ to $B$.
	To get an exact relation, the switching protocol is often assumed to be either very fast, as in free energy perturbation theory, or adiabatically slow, as in thermodynamic integration theory~\cite{Chipot2007}.   
	A major breakthrough was achieved with the introduction of the Jarzynski equality (JE)~\cite{Jarzynski1997a, Jarzynski1997b, Jarzynski2004},
	\begin{equation}
	\label{eq:JE}
	e^{- \beta \Delta F_0} = \langle e^{- \beta W_0^\text{in}}\rangle, 
	\end{equation}
whereby the free energy difference $\Delta F_0$ could be calculated from the exponential average of the work $W_0^\text{in}$ for any switching protocol {\it of arbitrary speed.}
	Here, the angular bracket denotes averaging over many repetitions of the switching protocol.
	The JE has lead to a plethora of new results in the context of nonequilibrium thermodynamics and statistical mechanics~\cite{Crooks1998, Crooks1999, Crooks2000, Hatano2001, Speck2005, Seifert2005, Esposito2010, Seifert2012}, and there have been recent advances in the thermodynamics of control~\cite{Sivak2012, Zulkowski2012,  Zulkowski2014, Zulkowski2015a, Zulkowski2015b}, prediction~\cite{Still2012}, self-replication~\cite{England2013}, and information processing~\cite{Sagawa2010, Horowitz2010, Sagawa2012, Mandal2012, Deffner2013, Ito2013, Horowitz2014, Parrondo2015}.

	Before the introduction of the JE, Bochkov and Kuzovlev had derived a similar relation (BKR)~\cite{Bochkov1977, Bochkov1981a, Bochkov1981b}, 
 	\begin{equation}
	\label{eq:BKR}
	1 = \langle e^{- \beta W_1^\text{ex}} \rangle, 
	\end{equation}
where, surprisingly, $W_1^\text{ex}$ is not equal to $(W_0^\text{in} - \Delta F_0)$, as one might expect from the JE [Eq.~(\ref{eq:JE})].
	(Note the definitions of $W_0^\text{in}$ and $W_1^\text{ex}$ given below in Eqs.~(\ref{eq:inclusive}) and (\ref{eq:exclusive}), respectively.)
	The apparent discrepancy between Eqs.~\ref{eq:JE} and \ref{eq:BKR} was resolved in~\cite{Jarzynski2007, Horowitz2007} by showing that these two relations correspond to two different conventions for defining internal energy, each leading to its own definition of work. 
	In fact, by considering the dynamics under two different time-dependent conservative forces, $f_0$ and $f_1$, Ref.~\cite{Horowitz2007} presented a unified expression 
	\begin{equation}
	\label{eq:Main_intro}
	e^{- \beta \Delta F_0} = \left \langle e^{- \beta \left(W_0^\text{in} + W_1^\text{ex}\right)} \right \rangle
	\end{equation} 
relating the free energy difference $\Delta F_0$ 
to the different measures of work $W_0^\text{in}$ and $W_1^\text{ex}$ for the two forces $f_0$ and $f_1$, respectively. 
The $0$ subscript on $\Delta F_0$ is meant to indicate that this free energy difference actually depends only on changes in $f_0$; $\Delta F_0$ is insensitive to changes in $f_1$.
We note that the detailed form of Eq.~\ref{eq:Main_intro} differs from the  corresponding formula presented in Ref.~\cite{Horowitz2007}, which assumes that the Hamiltonian is linear in $f_1$.

	We point out here the surprising fact that Eq.~(\ref{eq:Main_intro}) remains valid even if the system dynamics during the switching are not  related to the two Hamiltonians $\cal{H}_A$ and $\cal{H}_B$. 
		Unlike the JE, where the dynamics during switching are derived from a time-dependent Hamiltonian ${\cal H}(t): {\cal H}_A \rightarrow {\cal H}_B$, connecting the initial and final Hamiltonians, the combined JE and BKR formula [Eq.~(\ref{eq:Main_intro})] implies that the intermediate Hamiltonian ${\cal H}(t)$ can be independent of them. 
	More specifically, as long as the system is initiated in the equilibrium condition $\rho_A^\text{eq}$, the dynamics during the switching can be governed by any modified Hamiltonian ${\cal H}'(t) = {\cal H}(t) + {\cal H}_1(t)$, for arbitrary ${\cal H}_1(t)$. 
	To our knowledge, this aspect of the combined JE and BKR formula has not been appreciated before.

	Detailed studies have revealed the statistical quality of free energy estimation based on the JE~\cite{Sun2003, Oberhofer2005, Vaikuntanathan2008, Suarez2012}. 
	While slow driving protocols produce an effectively unbiased estimator, fast driving protocols induce far-from-equilibrium dynamics that often result in a bias.
	The convergence of a free energy estimator with respect to the number of independent samples is slow whenever the final phase space distribution of the system at the end of the protocol has a poor overlap with the final equilibrium distribution $\rho^\text{eq}_B$. 
	Several strategies have been employed to improve the convergence of the JE estimator for fast driving protocols, including modifying the system dynamics to enhance the overlap between the actual distribution of the system and $\rho^\text{eq}_B$, and employing bidirectional protocols. 
However, while some previous studies have exploited specific forms of non-Hamiltonian dynamics in order to improve free energy estimation, it has not been clear what the optimal strategy should be, and a general framework unifying and extending previous results has been lacking.

	The main contribution of this paper is to introduce a generalization of the combined JE and BKR (Eq.~[\ref{eq:Main_intro}]), given by Eq.~(\ref{eq:Main3}) below, that is compatible with non-Hamiltonian dynamics during switching. 
	In particular, we extend the former strategy to {\it arbitrary dynamics}, whereby the averaging in Eq.~\ref{eq:JE} is performed with the same initial equilibrium condition, $\rho^\text{eq}_A$, but the subsequent evolution is determined by potentially non-Hamiltonian dynamics, completely unrelated to the Hamiltonians ${\cal H}_{\cal A}$ and ${\cal H}_{\cal B}$.

	Even though our central result Eq.~\ref{eq:Main3} is valid for arbitrary dynamics, for optimal estimation, the dynamics must be tailored such that the actual distribution of the system at the end of the dynamics is the same as the equilibrium distribution $\rho^\text{eq}_B$.  
	Achieving this condition can make it possible to obtain accurate estimates after as few as one simulated transition.
	We derive the equation satisfied by such optimal modified dynamics and point out its relation to the so-called ``escorted dynamics"~\cite{Vaikuntanathan2008}. 
	We emphasize that, whereas the optimal estimation strategy for a broad class of systems involves the use of escorted dynamics, it is typically not the case that one can compute the specific form these dynamics take for interesting non-equilibrium systems, pointing to the need for a more general framework such as ours.

	The manuscript is organized as follows.
	In Sec. II we illustrate our derivation of the modified JE compatible with arbitrary dynamics (Eq.~\ref{eq:Main3}) for the simple case of the one-dimensional Langevin equation with only position-dependent forces. 
	We first consider the underdamped case and then we describe some important subtleties of the overdamped limit associated with different stochastic integration schemes. 
	In Sec. III, we give a general proof of Eq.~\ref{eq:Main3}.
	By construction, our general proof applies to situations involving many interacting Brownian particles.
	We show that, just like the JE, our result applies even when the dynamics take place in the absence of a thermal reservoir. 
	In Sec. IV we emphasize how there is a clear separation between the dynamics and the end-point Hamiltonians ${\cal H}_A$ and ${\cal H}_B$ in our new relation.
	In Sec. V, we derive an expression for optimal dynamics for free energy estimation based on Eq.~\ref{eq:Main3} and discuss its relation to the so-called escorted dynamics.
	We conclude by comparing our result to some recent studies on work-fluctuation relations in the absence of detailed balance.

	\section{One-dimensional example}

	Consider the isothermal dynamics of a particle of mass $m$ constrained to move on a circle of circumference $l$.  
	Let $x$ and $p$ denote its position and momentum, respectively, with the identification $x + l \equiv x$. 
	There are two forces from the heat reservoir, damping ($- \gamma p / m$) and noise ($\xi_t$), the latter having the following statistical properties:
	\begin{equation}
	\label{eq:Noise}
	\langle \xi_t \rangle = 0, \quad \langle \xi_t \xi_{t'} \rangle = \frac{2 \gamma}{\beta} \delta(t-t'). 
	\end{equation}
	In addition, we consider two other forces: $f_0(x; \lambda) = - \partial_x V(x; \lambda_0)$, derived from some potential $V(x; \lambda_0)$ with external parameter $\lambda_0$, and $f_1(x; \lambda_1)$, with external parameter $\lambda_1$, which is not necessarily derivable from a potential. 
	The dynamics of the particle are given by
	\begin{eqnarray}
	\label{eq:Langevin}
	\dot{x} & = & \frac{p}{m}, \nonumber \\
	\dot{p} & = &  f_0(x; \lambda_0) + f_1(x; \lambda_1) - \gamma \frac{p}{m} + \xi_t,
	\end{eqnarray}
where the dots over the variables $x$ and $p$ denote their time-derivatives.

	An equivalent way to describe the dynamics of the particle is via the Fokker-Planck equation for the phase-space probability density $\rho({\bf z}, t)$, with ${\bf z} = (x, p)$:
	\begin{equation}
	\label{eq:FP}
	\frac{\partial \rho}{\partial t}  =  \hat{\cal L} \rho = - {\boldsymbol \nabla}_{\bf z} \cdot {\bf J}, 
	\end{equation}
where $ {\bf J} = \left[\left(\frac{p}{m}, f_0 + f_1 - \gamma p - \frac{\gamma}{\beta} \partial_p \right) \rho \right]$ denotes the phase space probability current. 
	If the parameters $\{\lambda_0, \lambda_1\}$ are held fixed in time, it can be shown that any initial distribution $\rho({\bf z}, 0)$ relaxes to a unique stationary distribution $\rho^\text{s}({\bf z})$ with $\hat{\cal L} \rho^\text{s} = 0$~\cite{Risken1984}.  
	Note that, if the force $f_1$ is zero, the stationary distribution $\rho^\text{s}$ is the equilibrium distribution $\rho^\text{eq}({\bf z}; \lambda_0, \beta) \propto \exp{- \beta {\cal H}_{\lambda_0}}$,  with respect to the Hamiltonian ${\cal H}_{\lambda_0}({\bf z}) = p^2 / (2 m) + V(x; \lambda_0)$.

	Consider now a switching protocol specified by time-varying parameters $\{\lambda_0(t), \lambda_1(t)\}$ with $\lambda_0(0) = A$, $\lambda_0(\tau) = B$, and arbitrary $\lambda_1(t)$.
	Following Refs.~\cite{Jarzynski2007, Horowitz2007} we now introduce two different notions of work, \textit{inclusive} and \textit{exclusive}. 
	Inclusive work is applicable to only conservative forces while exclusive work is applicable to both conservative and nonconservative forces. 
	Inclusive work done by the conservative force $f_0(x; \lambda)$  for a given protocol $\{\lambda_0(t), \lambda_1(t)\}$ and over a trajectory $\{x(t), p(t)\}$ is
	\begin{equation}
	\label{eq:inclusive}
	W_0^\text{in} = \int \mathrm{d}t \, \dot{\lambda}_0(t) \frac{\partial V(x; \lambda_0)}{\partial \lambda_0}\bigg|_{x(t), \lambda_0(t)}. 
	\end{equation}
	The exclusive work done by the force $f_1(x; \lambda_1)$ is
	\begin{equation}
	\label{eq:exclusive}
	W_1^\text{ex} =   \int \mathrm{d}t \, \frac{p(t)}{m} \circ f_1(x(t); \lambda_1(t)), 
	\end{equation}
where the circle ($\circ$) on the right denotes Stratonovich multiplication~\cite{Sekimoto2010}.
	According to the Feynman-Kac theorem~\cite{Hummer2001, Vaikuntanathan2008}, the solution to the sink equation
	\begin{equation}
	\label{eq:Sink}
	\frac{\partial g}{\partial t} = \hat{\cal L} g - h g, 
	\end{equation}
with the initial condition $g({\bf z}, 0) = \rho^\text{eq}({\bf z}, \lambda_0, \beta)$ and arbitrary phase space function $h({\bf z}, t)$, is given by the average
	\begin{equation}
	\label{eq:FK}
	g({\bf z}, t) = \left \langle \delta({\bf z}(t) - {\bf z}) e^{-\int_0^t \mathrm{d}t' \, h(t')} \right \rangle
	\end{equation}
with $h(t) = h({\bf z}(t), t)$ and $\delta$ denoting the Dirac delta function. 
	While Eq.~(\ref{eq:FK}) is true of any $h$, if we consider the following particular form: 
	\begin{equation}
	\label{eq:hUD}
	h = \beta \dot{\lambda}_0 \partial_{\lambda_0} V + \beta \frac{p}{m} f_1,
	\end{equation}
its time-integral $\int \mathrm{d}t \, h$ is equivalent to the sum $W \equiv (W_0^\text{in} + W_1^\text{ex})$, in units of $1/ \beta$, despite the fact that we have not used Stratonovich multiplication in the last term of Eq.~(\ref{eq:hUD}) (Appendix A). 
	By direct substitution we can show that the unnormalized, time-dependent Boltzmann distribution 
	\begin{equation}
	\label{eq:unB}
	g({\bf z}, t) = \frac{1}{Z_0(A, \beta)} e^{- \beta {\cal H}_{\lambda_0(t)}({\bf z})},
	\end{equation}
is also a solution of Eq.~(\ref{eq:Sink}) for the special choice of $h$ given in Eq.~\ref{eq:hUD}.   
	Combining Eqs.~(\ref{eq:FK}) and~(\ref{eq:unB}) and integrating with respect to ${\bf z}$ at time $t = \tau$ we get the following equation, 
	\begin{equation}
	\label{eq:Main}
	e^{- \beta \Delta F_0} = \left \langle e^{- \beta \left(W_0^\text{in} + W_1^\text{ex}\right)} \right \rangle,
	\end{equation} 
which is a special case of our more general result, Eq.~(\ref{eq:Main3}), with $\Delta F_0 = F_0(B, \beta) - F_0(A, \beta)$. 
	Note that our approach --- applying the Feynman-Kac theorem to the original protocol, as opposed to applying a Crooks-like fluctuation theorem to forward and reverse trajectories as an intermediate step~\cite{Crooks1998, Crooks1999, Crooks2000, Seifert2012} --- gives a much quicker derivation of Eq.~(\ref{eq:Main}) compared to previous approaches that have been brought to bear on this one-dimensional example~\cite{Spinney2012}.	

	\subsection{Relation to fluctuation theorem for entropy production}

	There is a close connection between Eq.~(\ref{eq:Main}) and the integral fluctuation theorem for entropy production in the framework of stochastic thermodynamics~\cite{Seifert2012},
	\begin{equation}
	\label{eq:FT}
	\left\langle \frac{\rho_1({\bf z}(\tau))}{\rho_2({\bf z} (0))} e^{\beta Q} \right\rangle = 1,
	\end{equation} 
where $\rho_1$ and $\rho_2$ are any two normalized distributions and $Q$ is the heat supplied to the system, 
	\begin{equation}
	\label{eq:Heat}
	Q = \int \mathrm{d}t \, \frac{p(t)}{m} \circ \left[ - \gamma \frac{p(t)}{m} + \xi_t \right].
	\end{equation}
	$Q$ can also be thought of as the exclusive work done by the reservoir forces~\cite{Sekimoto2010}.
	If we consider the following forms for the $\rho_i$'s,
	\begin{equation}
	\label{eq:rhoss}
	\rho_1({\bf z}) = \rho^\text{eq}({\bf z}; B, \beta), \quad \rho_2({\bf z}) = \rho^\text{eq}({\bf z}; A, \beta),
	\end{equation}
then the fluctuation theorem~\ref{eq:FT} reduces to Eq.~(\ref{eq:Main}).
	One just needs to use conservation of energy at the level of each trajectory:
	\begin{equation}
	\label{eq:FirstLaw}
	\Delta {\cal H}_{\lambda_0} = W_0^\text{in} + W_1^\text{ex} + Q, 
	\end{equation} 
derivable from the Langevin Eq.~(\ref{eq:Langevin}).

	\subsection{Subtlety in the overdamped limit}

	Interestingly, our approach leads to a different integral fluctuation theorem than the entropy production fluctuation theorem in  the overdamped limit, described by the following dynamics\footnote{We have assumed uniform temperature and friction coefficient in the medium so that there is no anomalous contribution to the entropy production.}:  
	\begin{equation}
	\label{eq:LangevinOD}
	\dot{x} = \gamma^{-1} (f_0 + f_1) + \gamma^{-1} \xi_t,
	\end{equation}
often useful in molecular simulations.
	In this limit, the two definitions of work $W_0^\text{in}$ and $W_1^\text{ex}$ remain essentially unchanged; only $(p/m)$ in Eq.~(\ref{eq:exclusive}) needs to be replaced by $ \dot{x}$.
	From the fluctuation theorem for entropy production, Eq.~(\ref{eq:FT}), valid also in the overdamped limit, one can show the validity of Eq.~(\ref{eq:Main}) (Appendix B). 
	However, using the Feynman-Kac approach described above, one can derive the following relation (Appendix C)
	\begin{eqnarray}
	\label{eq:ODFK}
	e^{- \beta \Delta F^\text{c}_0} & = & \langle e^{- \int \mathrm{d}t \, h_\text{OD}} \rangle, \\
	\label{eq:hOD}
	h_\text{OD} & = & \beta \dot{\lambda}_0\, \partial_{\lambda_0} V - \frac{\beta}{\gamma} \left( f_0 f_1 + \frac{1}{\beta}\partial_x f_1 \right),
	\end{eqnarray}
where, unlike the underdamped scenario, the quantity in the exponent on the right of Eq.~\ref{eq:ODFK} is not equal to the sum $ \beta W = \beta (W_0^\text{in} + W_1^\text{ex})$.
	I.e., Eqs.~\ref{eq:Main} and \ref{eq:ODFK} are not the same.  
	Note that the free energy $F_0^\text{c}(\lambda_0, \beta)$ in this context is the configurational free energy
	\begin{equation}
	\label{eq:FreeEnConfig}
	e^{-\beta F^\text{c}_0(A, \beta)} = Z_0^\text{c}(\lambda_0, \beta)=  \int \mathrm{d}x \, e^{- \beta V(x; \lambda_0)}.
	\end{equation}
	This duality of integral fluctuation theorems, observable only in the overdamped limit, has been reported before~\cite{Liu2014}, but only in the context of the Bochkov-Kuzovlev relation, Eq.~\ref{eq:BKR}.
	In contrast, our treatment proves the existence of this duality in the broader context we consider here.
	Our preliminary investigation suggests that the duality stems from the effects of the transformation $f_1 \rightarrow - f_1$ on the probabilities of trajectories. 
	We reserve the detailed investigation for a future study.

	We note that, from the perspective of numerical computation of $\Delta F_0^\text{c}$, it is advantageous to use Eq.~(\ref{eq:ODFK}) instead of Eq.~(\ref{eq:Main}), because the latter involves cancellation of large integrals, which is numerically costly. 
	The requirement of the cancellation is easily seen via the Ito representation of the sum $W = (W_0^\text{in} + W_1^\text{ex})$: 
	\begin{equation}
	\label{eq:WOD}
	W = \int \mathrm{d}t \, \left[\dot{\lambda}_0 \partial_{\lambda_0} V + \frac{1}{\gamma} \left( f_0 f_1 + \frac{1}{\beta} \partial_x f_1 + f_1^2 + \xi_t f_1 \right) \right]. 
	\end{equation}
	The fourth term on the right is nonnegative and its integral grows with time. 
	Convergence of the right hand side of Eq.~(\ref{eq:Main}) can be achieved only if this growing integral is canceled at each power.

 	\section{Fluctuation theorem valid for general dynamics}
	\label{sec:General}

	We now derive a new fluctuation theorem that is valid even for general non-Hamiltonian dynamics (Eq.~(\ref{eq:Main3})). 
	Consider a system described by the phase space coordinates ${\bf z} = \{{\bf x}, {\bf p}\}$ and the Hamiltonian 
	\begin{equation}
	\label{eq:Hamiltonian}
	{\cal H}_{{\boldsymbol \lambda}_0}({\bf z}) = \frac{p^2}{2 m} + V({\bf x}; {\boldsymbol \lambda}_0),
	\end{equation} 
where $p$ is the magnitude of momentum ${\bf p}$ and ${\boldsymbol \lambda}_0 = \{\lambda_{01}, \lambda_{02}, \ldots \}$ is a set of external parameters. 
	For a system with many particles, all of mass $m$, we have ${\bf p} = ({\bf p}_1, {\bf p}_2, \ldots)$ and ${\bf x} = ({\bf x}_1, {\bf x}_2, \ldots)$ for particles 1, 2, and so on. 
	If we couple the system to a thermal reservoir of inverse temperature $\beta$, quite generally, we can write down the equation to motion to be of the following form\footnote{One can use different mass values $m_i$ for different particles without changing the final result.}:
	\begin{eqnarray}
	\label{eq:EoM}
 	\dot{\bf x} & = & \frac{\bf p}{m}, \nonumber \\
	\dot{\bf p} & = & - {\boldsymbol \nabla}_{\bf x} V({\bf x}; {\boldsymbol \lambda}_0) - \Gamma \frac{{\bf p}}{m} + \Xi_t,
	\end{eqnarray}
with $\langle \Xi_t \rangle = 0$ and $\langle \Xi_t \Xi^{\text T}_{t'}\rangle = 2 (\Gamma / \beta) \delta(t-t')$. 
	Here, $\Gamma$ is a positive definite matrix denoting the damping coefficient matrix and $\Xi$ is the noise vector.  
	The phase space distribution $\rho({\bf z}, t)$ evolves according to the Fokker-Planck equation
	\begin{equation}
	\label{eq:FP2}
	\frac{\partial \rho} {\partial t} = \hat{{\cal L}}_0 \rho = - {\boldsymbol \nabla}_{\bf z} \cdot {\bf J},
	\end{equation}
with ${\bf J} = \left[\left( {\bf p}/m, - {\boldsymbol \nabla}_{\bf x} V({\bf x}; {\boldsymbol \lambda}_0) - \Gamma {\bf p} - (\Gamma/ \beta) {\boldsymbol \nabla}_{\bf p} \right) \rho \right]$. 
	As in the one-dimensional example, the asymptotic solution is the Boltzmann distribution: 
	\begin{eqnarray}
	\label{eq:Eql2}
	\rho^\text{eq}({\bf z}; {\boldsymbol \lambda}_0, \beta) & = & e^{- \beta \left[{\cal H}_{{\boldsymbol \lambda}_0} - F_0({{\boldsymbol \lambda}_0}, \beta)\right]}, \\
	\label{eq:FreeEn3}
	e^{ -\beta F_0({{\boldsymbol \lambda}_0}, \beta)} & = & Z_0({\boldsymbol \lambda}_0, \beta) = {\int \mathrm{d}{\bf z} \, e^{- \beta {\cal H}_{{\boldsymbol \lambda}_0}}}.
	\end{eqnarray} 
	We now add an arbitrary phase space velocity vector 
	\begin{equation}
	\label{eq:v1}
	{\bf v}_1 = \left[ {\bf f}_{1 {\bf x}}({\bf z}; {\boldsymbol \lambda}_1), {\bf f}_{1 {\bf p}}({\bf z}; {\boldsymbol \lambda}_1) \right]
	\end{equation} 
to the dynamics of the system with external parameters ${\boldsymbol\lambda}_1 = \{\lambda_{11}, \lambda_{12}, \ldots\}$, leading to the following modified dynamics
	\begin{eqnarray}
	\label{eq:EoM2}
 	\dot{\bf x} & = & \frac{\bf p}{m} + {\bf f}_{1{\bf x}}, \nonumber \\
	\dot{\bf p} & = & {\bf f}_0+ {\bf f}_{1 {\bf p}} - \Gamma \frac{{\bf p}}{m} + \Xi_t, 
	\end{eqnarray}
where we have defined 
	\begin{equation}
	{\bf f}_0 = - {\boldsymbol \nabla}_{\bf x} V({\bf x}; {\boldsymbol \lambda}_0).
	\end{equation} 
	Such additional phase space velocity vectors arise in many different contexts: 
	(i) for velocity dependent feedback control, with ${\bf v}_1 = ({\bf 0}, - \Gamma {\bf p}/m)$ for some stable matrix $\Gamma$~\cite{Kim2004, Kim2007}; 
	(ii) for self-propelled active particles, with ${\bf v}_1 = \left[ {\bf 0}, {\bf F}({\bf p}/m) + {\bf f}(t)\right]$ for some odd function ${\bf F}$ and a generic function ${\bf f}$~\cite{Ganguly2013}; and 
	(iii) for escorted, simulation dynamics, with ${\bf v}_1 = \sum_i \dot{\lambda}_{0i} {\bf u}_i({\bf z}; {\boldsymbol \lambda}_0)$ for arbitrary, continuous phase space vector fields ${\bf u}_i$~\cite{Miller2000, Vaikuntanathan2008}.  
	Note that ${\bf v}_1$ can arise either from real physical forces, as in cases (i) and (ii) above, or from artificial dynamics intended to facilitate computer simulation and sampling of a system, as in case (iii). 
	Note also that ${\bf v}_1$ does not generally follow from any Hamiltonian. 
	The addition of ${\bf v}_1$ leads to a modified Fokker-Planck operator
	\begin{equation}
	\label{eq:mFP}
	\hat{{\cal L}} = \hat{{\cal L}}_0 + \hat{{\cal L}}_1, \quad \hat{{\cal L}}_1 \rho = - {\boldsymbol \nabla}_{\bf z} \cdot ({\bf v}_1 \rho),
	\end{equation} 
leading to a modified stationary  distribution $\rho^\text{s}({\bf z})$, $\hat{{\cal L}} \rho^\text{s} = 0$. 
	Even when ${\bf v}_1$ has a physical origin, i.e., it is of the form ${\bf v}_1 = \left[ {\bf 0}, {\bf f}_{1 {\bf p}}({\bf z}; {\boldsymbol \lambda}_1) \right]$ for some physical force ${\bf f}_{1 {\bf p}}({\bf z}; {\boldsymbol \lambda}_1)$, the stationary distribution $\rho^\text{s}$ may be unknown if the force is not derivable from a potential.

	Consider now initiating the system at the equilibrium distribution $\rho^\text{eq}({\bf z}; {\boldsymbol \lambda}_0, \beta)$ and driving the system according to some protocol ${\boldsymbol \lambda}(t) = \{{\boldsymbol \lambda}_0(t), {\boldsymbol \lambda}_1 (t)\}$, as ${\boldsymbol \lambda}_0$ varies from ${\bf A}$ to ${\bf B}$. 
	We wish to calculate the free energy difference $\Delta F_0 = \left[ F_0({\bf B}, \beta) - F_0({\bf A}, \beta)\right]$.
	At any point along any trajectory ${\bf z}(t)$, the inclusive power by the original, conservative forces is given by 
	\begin{equation}
	\label{eq:Win2}
	\dot{W}_0^\text{in} = \dot{{\boldsymbol \lambda}}_0 \cdot {\boldsymbol \nabla}_{{\boldsymbol \lambda}_0} V({\bf z}(t);{{\boldsymbol \lambda}_0}),
	 \end{equation} 
and the exclusive power by the additional force ${\mathbf f}_{1 {\bf p}}({\bf z}(t); {\boldsymbol \lambda}_1)$ is given by 
	\begin{equation}
	\label{eq:Wex2}
	\dot{W}_1^\text{ex} = {\mathbf f}_{1 {\bf p}}({\bf z}(t); {\boldsymbol \lambda}_1) \circ {\bf p}(t)/m.
	\end{equation} 
	However, unlike the one-dimensional case, the average of the exponential of minus the sum $\beta (W_0^\text{in} + W_1^\text{ex})$ does not give the free energy change $\Delta F_0$ in this general case. 
	We need to consider additional terms.
	To see this, let us begin with the unnormalized distribution 
	\begin{equation}
	\label{eq:unB2}
	g({\bf z}, t) = \frac{1}{Z_0({\bf A}, \beta)} e^{- \beta H_{{\boldsymbol \lambda}_0(t)}({\bf z})}.
	\end{equation}
	By substituting it into the sink equation
	\begin{equation}
 	\label{eq:Sink2}
	\frac{\partial g}{\partial t} = \hat{\cal L} g - h g,
	\end{equation}
and requiring that the right hand side of Eq.~(\ref{eq:unB2}) is a solution of Eq.~(\ref{eq:Sink2}), we obtain the following expression for $h$:
	\begin{equation}
	\label{eq:hUD2}
	h = \beta \left(\dot{W}_0^\text{in} + \dot{W}_1^\text{ex}\right) - {\boldsymbol \nabla}_{\bf z} \cdot {\bf v}_1 + \beta {\bf f}_{1 {\bf x}} \cdot {\boldsymbol \nabla}_{\bf x} V.
	\end{equation}
	Finally, applying the Feynman-Kac theorem to the sink equation~(\ref{eq:Sink2}) with $h$ as defined in Eq.~(\ref{eq:hUD2}), and combining with Eq.~(\ref{eq:unB2}), we get 
	\begin{equation}
	\label{eq:Main3}
	e^{- \beta \Delta F_0} = \langle e^{- \beta \left(W_0^\text{in} + W_1^\text{ex}\right) + \int \mathrm{d} t \, \left( \nabla_{\bf z} \cdot {\bf v}_{1} - \beta {\bf f}_{1 {\bf x}} \cdot {\boldsymbol \nabla}_{\bf x} V \right) } \rangle,
	\end{equation}
	which is our main result.

	Eq.~(\ref{eq:Main3}) generalizes Eq.~(\ref{eq:Main}) to the situation where the additional field ${\bf v}_1 = ({\bf f}_{1 {\bf x}}, {\bf f}_{1 {\bf p}})$ need not involve solely position-dependent forces.  
	In particular, Eq.~(\ref{eq:Main3}) includes as a special case Brownian dynamics under electromagnetic forces~\cite{Pradhan2010a}. 
	In this case, surprisingly, the integral in the exponent on the right of Eq.~(\ref{eq:Main3}) drops out leading to Eq.~(\ref{eq:Main}), as observed in~\cite{Pradhan2010a}. 
	In a general scenario, both terms in the integral are non-vanishing. 
	The first term, $\int \mathrm{d}t \, {\boldsymbol \nabla}_{\bf z} \cdot {\bf v}_1$, accounts for the phase space contraction if the additional velocity term ${\bf v}_1$ is dissipative. 
	In the continuous feedback literature, this term has been referred to as entropic pumping~\cite{Kim2007}. 
	The meaning of the second additional term, $ - \int \mathrm{d}t \, {\bf f}_{1 {\bf x}} \cdot {\boldsymbol \nabla}_{\bf x} V$, is less transparent, probably because such terms do not appear to have any physical origin, though they can arise in artificial, simulation dynamics~\cite{Vaikuntanathan2008}.

	\subsection{Thermally isolated dynamics}
	\label{sec:Isolated}

	Just like the original JE, a feature of Eq.~(\ref{eq:Main3}) is that it is valid even when the dynamics during the switching are thermally isolated~\cite{Jarzynski1997a}.	
	In this case, the system is initiated in the same equilibrium distribution, $\rho^\text{eq}({\bf z}; {\bf A}, \beta)$, but the subsequent evolution does not involve the reservoir terms in Eq.~\ref{eq:EoM2}, i.e, the system evolves according to the following dynamics:
	\begin{equation}
	\label{eq:EoM3}
 	\dot{\bf x} =  \frac{\bf p}{m} + {\bf f}_{1{\bf x}}, \quad \dot{\bf p}  =  {\bf f}_0+ {\bf f}_{1 {\bf p}}.
	\end{equation}
	The proof based on Feynman-Kac theorem still applies with a modified Fokker-Planck operator~\cite{Vaikuntanathan2008}. 
	However, the following derivation provides more insight. 
	We evaluate the average 
$\langle \exp{(- \int \mathrm{d}t \, h)} \rangle$ (with $h$ given by Eq.~\ref{eq:hUD2}) over many repetitions of the protocol ${\boldsymbol \lambda}(t) = \{{\boldsymbol \lambda}_0(t), {\boldsymbol \lambda}_1 (t)\}$.  
	We can rewrite the integral $ (1/\beta) \int \mathrm{d}t \, h$ along any phase space trajectory ${\bf z}(t)$ as 
	\begin{align}
	\label{eq:Sh}
	& \frac{1}{\beta} \int \mathrm{d}t \, h \nonumber \\
	= & \int \mathrm{d}t \, \left[ \left({\boldsymbol \lambda}_0 \cdot {\boldsymbol \nabla}_{{\boldsymbol \lambda}_0} + {\bf f}_{1 {\bf p}} \cdot {\boldsymbol \nabla}_{\bf p} + {\bf f}_{1 {\bf x}} \cdot {\boldsymbol \nabla}_{\bf x} \right) {\cal H}_{{\boldsymbol \lambda}_0} - (1/\beta){\boldsymbol \nabla}_{\bf z} \cdot {\bf v}_1 \right] \nonumber \\
	 = & \int \mathrm{d}t \, \left( \frac{d}{dt} {\cal H}_{{\boldsymbol \lambda}_0} -  (1/\beta){\boldsymbol \nabla}_{\bf z} \cdot \dot{\bf z}\right) \nonumber \\
	 = & {\cal H}_{\bf B}({\bf z}(\tau)) - {\cal H}_{\bf A}({\bf z}(0)) - (1/\beta) \int \mathrm{d}t\, {\boldsymbol \nabla}_{\bf z} \cdot \dot{\bf z},
	\end{align}
where we have used Eqs.~\ref{eq:hUD2}, \ref{eq:Win2}, and \ref{eq:Wex2} in the first line and Eq.~\ref{eq:EoM3} in the second line. 
	Note that, because the evolution of the system takes place in the absence of a thermal reservoir, the evolution is deterministic --- if we know the initial phase space coordinate ${\bf z}(0)$, we know the future trajectory ${\bf z}(t > 0)$ for any given protocol $\{{\boldsymbol \lambda}_0 (t), {\boldsymbol \lambda}_1(t) \}$. 	
	As a result, the integral $ (1/\beta) \int \mathrm{d}t \, h$ can be treated as a function of just ${\bf z}(0)$. 
	In particular, we can rewrite the average $\langle \exp{(- \int \mathrm{d}t \, h)} \rangle$ as 
	\begin{equation}
	\label{eq:SMain0}
	\langle e^{- \int \mathrm{d}t \, h} \rangle = \int \mathrm{d}{\bf z}(0) \rho^\text{eq}({\bf z}(0); {\bf A}, \beta) \, e^{- \int \mathrm{d}t \, h }.
	\end{equation}
	We can simplify Eq.~\ref{eq:SMain0} further:
	\begin{subequations}
	\begin{eqnarray}
	\label{eq:SMain}
	\langle e^{- \int \mathrm{d}t \, h} \rangle & = & \int \mathrm{d}{\bf z}(0) \, \frac{e^{- \beta {\cal H}_{\bf A}({\bf z}(0)) - \int \mathrm{d}t \, h }}{Z_0({\bf A}, \beta)} \\
	\label{eq:SMainb}
	& = & \int \mathrm{d}{\bf z}(0) \frac{e^{- \beta {\cal H}_{\bf B}({\bf z}(\tau))}}{Z_0({\bf A}, \beta)} e^{\int \mathrm{d}t \, {\boldsymbol \nabla}_{\bf z} \cdot \dot{\bf z}} \\
	\label{eq:SMainc}
	& = & \int \mathrm{d}{\bf z}(\tau) \frac{e^{- \beta {\cal H}_{\bf B}({\bf z}(\tau))}}{Z_0({\bf A}, \beta)} \\
	\label{eq:SMaind}
	& = & \frac{Z_0({\bf B}; \beta)}{Z_0({\bf A}, \beta)} \\
	& = & e^{- \beta \left[ F_0({\bf B}, \beta) - F_0({\bf A}, \beta)\right]}. 
	\end{eqnarray}
	\end{subequations}
	In line~\ref{eq:SMain}, we have rewritten Eq.~\ref{eq:SMain0}; in line~\ref{eq:SMainb}, we have used Eq.~\ref{eq:Sh} for the integral $\int \mathrm{d}t \, h$; in line \ref{eq:SMainc}, we have used the fact that the Jacobian $|\partial {\bf z}(\tau) / \partial {\bf z}(0)|$ is given by $\exp{\left(\int \mathrm{d}t \, {\boldsymbol \nabla}_{\bf z} \cdot \dot{\bf z}\right)}$; and in the last two lines we have used the definitions in Eq.~\ref{eq:FreeEn3}. 
	This completes our alternate derivation of Eq.~\ref{eq:Main3} for thermally isolated dynamics.

	\section{Arbitrary dynamics}

	We want to emphasize that the dynamics during the protocol ${\boldsymbol \lambda}_0(t)$ can be completely independent of ${\cal H}_{{\boldsymbol \lambda}_0}$ and we will still have Eq.~\ref{eq:Main3}.
	Consider an additional field ${\bf v}_1$ of the form
	\begin{equation}
	\label{eq:Independence}
	{\bf f}_{1 {\bf x}} = - \frac{{\bf p}}{m} + \tilde{{\bf f}}_{1 {\bf x}}, \quad {\bf f}_{1 {\bf p}} =  \frac{ \partial V({\bf x}; {\boldsymbol \lambda}_0)}{\partial {\bf x}} + \tilde{{\bf f}}_{1 {\bf p}}, 
	\end{equation}  
for any given ${\boldsymbol \lambda}_0(t)$ and arbitrary $\tilde{{\bf v}}_1 = (\tilde{{\bf f}}_{1 {\bf x}}, \tilde{{\bf f}}_{1 {\bf p}})$. 
	The system then evolves according to 
	\begin{eqnarray}
	\label{eq:EoMInd}
	\dot{\bf x} & = & \tilde{\bf f}_{1 {\bf x}}, \nonumber \\
	\dot{\bf p} & = & \tilde{\bf f}_{1 {\bf p}} - \Gamma \frac{\bf p}{m} + \Xi_t, 
	\end{eqnarray}
without any term related to the Hamiltonian ${\cal H}_{{\boldsymbol \lambda}_0(t)}$, and yet we will still recover the free energy difference $\Delta F_0$ from Eq.~\ref{eq:Main3}.
	(The reservoir terms $\Gamma \frac{\bf p}{m}$ and $\Xi_t$ will be missing in context of thermally isolated evolution of Sec.~\ref{sec:Isolated}.) 
	This indicates an interplay between the dynamics and the quantity to average on the right hand side of Eq.~\ref{eq:Main3} which keeps the left hand side intact. 
	This level of flexibility in choosing the dynamics seems not to have been appreciated before.
	Another benefit of the current approach is that we can quickly derive Eq.~(\ref{eq:Main3}) without going into detailed considerations of path integrals and conjugate processes~\cite{Horowitz2007, Spinney2012, Seifert2012}. 

	\section{Optimal dynamics}

	The dissociation between the dynamics and the initial and final equilibrium states promises to facilitate a range of techniques for free energy estimation in a single, universal expression.
	Indeed such an instance has been seen before\cite{Vaikuntanathan2008} for a special class of additional phase space velocity vector ${\bf v}_1$ referred to as escorted dynamics.
	In the presence of a single time-dependent parameter $\lambda_0(t)$, the following form of ${\bf v}_1$ was chosen,
	\begin{equation}
	\label{eq:ED}
	{\bf v}_1 = \dot{\lambda}_0 {\bf u}({\bf z}; { \lambda}_0),
	\end{equation} 
 and it was shown that with an appropriate choice of ${\bf u}({\bf z}; { \lambda}_0)$ it is possible to vastly improve the statistical quality of the free energy estimator based on Jarzynski-like relations. 
	The essential idea behind choosing the appropriate dynamics was to ensure that the distribution of the system under the modified dynamics, $\rho({\bf z}; t)$, evolves close to the time-dependent equilibrium distribution $\rho^\text{eq}({\bf z}; {\lambda}_0(t))$. 
	In fact, an exact equation was proposed for the optimal choice of ${\bf u}({\bf z}; \lambda_0)$ by requiring that the time-dependent distribution is exactly the same as $\rho^\text{eq}({\bf z}; {\lambda}_0(t))$. 
	In this section, we consider the case of more than one time-dependent parameter.
	We see that the optimal dynamics for free energy estimation can be recast in a generalized version of Eq.~\ref{eq:ED} in an extremely general setting.

	In order to obtain the optimal dynamics, such that a single instantiation is sufficient to yield an accurate estimate of the free energy difference, we can impose the condition that  $\rho^\text{eq}({\bf z}; {\boldsymbol \lambda}_0(t))$ is a solution of the modified Fokker-Planck equation 
	\begin{equation}
	\label{eq:FP3}
	\frac{\partial \rho} {\partial t} = \hat{{\cal L}} \rho,
	\end{equation}
(see Eqs.~\ref{eq:FP2} and \ref{eq:mFP}) then, after some algebra, we get the following equation\footnote{As before, we are assuming isothermal dynamics. However, we might also consider time-dependent temperature by considering $\beta$ as a parameter in the set ${\boldsymbol \lambda}_0$.} for optimal ${\bf v}_1({\bf z}, t)$, denoted by ${\bf v}_1^*$:
	\begin{equation}
	\label{eq:v1Optimal}
	{\boldsymbol \nabla}_{\bf z} \cdot {\bf v}_1^* - \beta {\bf v}_1^* \cdot {\boldsymbol \nabla}_{\bf z} {\cal H}_{{\boldsymbol \lambda}_0} = - \beta \dot{\boldsymbol \lambda}_0(t) \cdot {\boldsymbol \nabla}_{{\boldsymbol \lambda}_0} \left[ {\cal H}_{{\boldsymbol \lambda}_0} - F_0({\boldsymbol \lambda}_0; \beta)\right].
	\end{equation}
	In this case, we can obtain a solution of a form similar to Eq.~\ref{eq:ED} by considering an additional phase space velocity field ${\bf v}_{1i}^*$ for each external parameter ${\lambda}_{0i}$, {\em i.e.}, by considering
	\begin{equation}
	\label{eq:ED2}
	{\bf v}_1^* = \sum_i \dot{\lambda}_{0i} {\bf u}_i^*({\bf z}; {\boldsymbol \lambda}_0),
	\end{equation}
with each additional field ${\bf u}_i^*$ now satisfying the following equation (under the assumption that all of the $\dot{\lambda}_{0i}(t)$ are non-zero)
	\begin{equation}
	\label{eq:u1Optimal}
	{\boldsymbol \nabla}_{\bf z} \cdot {\bf u}_{i}^* - \beta {\bf u}_{i}^* \cdot {\boldsymbol \nabla}_{\bf z} {\cal H}_{{\boldsymbol \lambda}_0} = - \beta  \partial_{\lambda_{0i}} \left[ {\cal H}_{{\boldsymbol \lambda}_0} - F_0({\boldsymbol \lambda}_0; \beta)\right].
	\end{equation}
	Equation~\ref{eq:u1Optimal} can be simplified further due to the form of the Hamiltonian given in Eq.~\ref{eq:Hamiltonian}.
	Consider the notation ${\bf u}_1^* = ({\bf u}_{1 {\bf x}}^*, {\bf u}_{1 {\bf p}}^*)$. 
	We can consistently assume that the momentum components, ${\bf u}_{i {\bf p}}^*$, are zero and get the following equation for ${\bf u}_{i{\bf x}}^*({\bf x}; {\boldsymbol \lambda})$:
	\begin{equation}
	\label{eq:uixOptimal}
	{\boldsymbol \nabla}_{\bf x} \cdot {\bf u}_{i{\bf x}}^* - \beta {\bf u}_{i{\bf x}}^* \cdot {\boldsymbol \nabla}_{\bf x}  V({\bf x};{{\boldsymbol \lambda}_0}) = - \beta \partial_{\lambda_{0i}} \left( V - F_0\right).
	\end{equation}

	As may be seen from Eqs.~\ref{eq:v1Optimal}, \ref{eq:u1Optimal}, and \ref{eq:uixOptimal}, equations for the optimal dynamics are complicated and they even involve the free energy itself that we are trying to calculate. 
	Clearly, it is extremely unlikely that one could derive the optimal dynamics in all but the simplest cases.
	Nonetheless, as for the case of escorted dynamics, the current approach provides insight into how to choose ${\bf v}_1$ such that free energy estimation is enhanced. 
	The fact that escorted dynamics are already sufficiently powerful to provide the optimal dynamics for estimation might seem to suggest that there would be no practical benefit to developing tools compatible with a broader class of dynamics.  
	However, note that we would already need to know the free energy difference, the very quantity we are trying to estimate, in order to solve Eq.~(\ref{eq:v1Optimal}) in nearly any physical system of interest, which is why our more general relation Eq.~(\ref{eq:Main3}) promises to be useful for efficient free energy estimation.

	\section{Discussion}

	The current work should be contrasted with the studies described in Refs.~\cite{Chernyak2006, Imparato2006, Yin2006, Kurchan2007, Pradhan2010b, deOliveira2011, Spinney2012, Ganguly2013, Ichiki2013, Tang2014, Ohzeki2014}. 
	In~\cite{Ganguly2013} only velocity dependent additional forces were considered whereas in Refs.~\cite{Chernyak2006, Imparato2006, Kurchan2007, Spinney2012} only position dependent forces were considered.   
	In~\cite{Pradhan2010b, deOliveira2011}, only those additional forces were considered for which the steady state distribution was of the Boltzmann form, which is not the case in the current work. 
	In~\cite{Yin2006, Tang2014}, the authors started with a generic Langevin equation, not derived from a Hamiltonian, and tried to build a thermodynamic theory for the dynamics. 
	Their approach was based on the decomposition of their abstract dynamics into reversible and irreversible components. 
	Our approach is complementary to theirs as we start from a given Hamiltonian and then add forces that may be nonconservative. 
Finally, in~\cite{Ohzeki2014} the authors propose a speed up of the calculation of free energy differences by utilizing the fact that violation of detailed balance can be used to accelerate the relaxation to steady states~\cite{Ichiki2013}.
	Given that Eq.~(\ref{eq:Main3}) is valid even in the absence of detailed balance, it will be interesting to investigate whether our new relation will lead to yet faster algorithms for calculating changes in free energy. 
	
Another interesting direction for future research concerns the further generalization of our results to include the determination of free energy profiles along reaction coordinates, as opposed to the free energy difference just between two given equilibrium states. 
	Also, it remains to be seen to what extent the framework of bidirectional protocols developed for the Jarzynski relation and its generalization by Hummer and Szabo~\cite{Hummer2001} may be developed for our new relation~\cite{Shirts2003, Minh2008}.   
	This could yield even more efficient approaches for calculating changes in free energy from limited data or simulations.

	In addition to the benefits of more efficient free energy estimation techniques for physical systems that actually obey Hamiltonian dynamics, the importance of developing a general framework compatible with non-Hamiltonian dynamics is highlighted by a recent example \cite{Argun2016} of an ``active matter," non-equilibrium system that cannot be handled by the JE, the combined JE and BKR, or any other previous generalizations that we are aware of.
	This particular system consists of a colloidal particle in contact with an ``active bath" containing bacteria that themselves dissipate heat and create microscopic structure in the solution, but other types of active matter systems have been observed~\cite{Ramaswamy2010, Marcetti2012}.
	We hope and expect that our framework will facilitate the study of active matter systems, as well as systems subject to continuous feedback, which to our knowledge are not correctly described by any previous generalization of the JE or the combined JE and BKR.

	\acknowledgements
	We acknowledge many useful discussions with C. Jarzynski, G. E. Crooks, F. van Wijland, S. Vaikuntanathan, A. Dhar, D. A. Sivak and P. R. Zulkowski. M.\ R.\ D.\ is grateful for support from the National Science Foundation through Grant No. IIS-1219199.
	This material is based upon work supported in part by the U.S. Army Research Laboratory and the U.S. Army Research Office under contract number W911NF-13-1-0390.

	\appendix

	\section{Irrelevance of Stratonovich scheme in underdamped dynamics}

	Here we show the equivalence of the following stochastic integrals in the context of underdamped Langevin dynamics:
	\begin{eqnarray}
	\label{eq:SStratonovich}
	W_0^\text{in} + W_1^\text{ex} & = & \int \mathrm{d}t \, \left( \dot{\lambda}_0 \partial_{\lambda_0} V + f_1 \circ \frac{p}{m} \right), \\
	\label{eq:SIto}
	\frac{1}{\beta} \int \mathrm{d}t \, h & = & \int \mathrm{d}t \, \left( \dot{\lambda}_0 \partial_{\lambda_0}V + f_1 \frac{p}{m} \right). 
	\end{eqnarray}
	The circle ($\circ$) on the right of Eq.~(\ref{eq:SStratonovich}) denotes Stratonovich multiplication~\cite{Sekimoto2010}. 
	The Ito representation of integral~(\ref{eq:SStratonovich}) can be obtained through the following steps:
	\begin{subequations}
	\begin{align}
	 & \int \mathrm{d}t \, \left( \dot{\lambda}_0 \partial_{\lambda_0} V + f_1 \circ \frac{p}{m} \right) \nonumber \\
	\label{eq:SItoStra}
	= & \int \mathrm{d}t \, \left[ \dot{\lambda}_0 \partial_{\lambda_0} V +  \frac{f_1(t) + f_1(t + dt)}{2} \frac{p}{m} \right] \\
	\label{eq:SItoStrb}
	= & \int \mathrm{d}t \, \left[ \dot{\lambda}_0 \partial_{\lambda_0} V +  f_1(t) \frac{p}{m} + \frac{f_1(t + dt) - f_1(t)}{2} \frac{p}{m} \right] \\
	\label{eq:SItoStrc}
	\approx & \int \mathrm{d}t \, \left( \dot{\lambda}_0 \partial_{\lambda_0}V + f_1(t) \frac{p}{m} \right) \nonumber \\
	& + \int \mathrm{d}t \, \frac{\left[ \left( \partial_x f_1 \right)_t dx + \left( \partial_{\lambda_1} f_1 \right)_t d \lambda_1 \right]}{2} \frac{p}{m}.
	\end{align}
	\end{subequations}
In line~\ref{eq:SItoStra} we have used the definition of Stratonovich integration; in line~\ref{eq:SItoStrb} we have simply rearranged terms; and in line~\ref{eq:SItoStrc} we have used the Taylor expansion of $f_1(t + dt)$ to first order in $dt$; $f_1$ is a function of $x$ and $\lambda_1$, both of which depend on time.  
	We have neglected the higher order terms in the Taylor expansion for the reasons given below.

	The first integral in line~\ref{eq:SItoStrb} is equal to the integral in Eq.~\ref{eq:SIto}.  
	In the following, we argue that the other terms in line~\ref{eq:SItoStrc} are negligible.  
	According to the underdamped Langevin equation [Eq.~\ref{eq:Langevin} of the main text], the differential $dx$ varies as $dt$. 
	If we assume the given protocol $\lambda_1(t)$ to be continuous and smooth, the differential $d \lambda_1$ also varies as $dt$. 
	We may conclude that the whole integrand in the last integral of line~\ref{eq:SItoStrb} varies as $dt$, and therefore the integral is zero. 
	The higher order terms in the Taylor expansion of line~\ref{eq:SItoStra} vary as higher powers of $dt$, and their contributions are zero as well.
	Combining Eqs.~\ref{eq:SIto}, \ref{eq:SItoStra} and \ref{eq:SItoStrb}, we find that the two integrals in Eqs.~\ref{eq:SStratonovich} and \ref{eq:SIto} are equivalent.

	\section{Derivation of Eq.~\ref{eq:Main} for one-dimensional overdamped dynamics}

	Using the overdamped Langevin equation [Eq.~\ref{eq:LangevinOD} of the main text], we can derive the following form of the first law of thermodynamics valid at the level of each realization $x(t)$: 
	\begin{equation}
	\label{eq:SFirstLaw}
	\Delta E = Q + W_0^\text{in} + W_1^\text{ex},
	\end{equation}	
where the energy $E$ is equal to the potential energy $V$; heat $Q$ is given by Eq.~\ref{eq:Heat} of the main text, with $p/m$ being replaced by $\dot{x}$; and the two types of work $W_0^\text{in}$ and $W_1^\text{ex}$ are given by Eqs.~\ref{eq:inclusive} and \ref{eq:exclusive} of the main text, respectively, again with the replacement $p/m \rightarrow \dot{x}$. 
	If we use Eq.~\ref{eq:SFirstLaw} in Eq.~\ref{eq:FT} of the main text, with $\rho_1$ and $\rho_2$ having the following forms,
	\begin{equation}
	\label{eq:SBoltz}
	\rho_1 \propto e^{- \beta V(x; B)} \quad , \quad \rho_2 \propto e^{- \beta V(x; A)},
	\end{equation}
after some cancellation of terms, we arrive at Eq.~\ref{eq:Main}.

	\section{Derivation of Eq.~\ref{eq:ODFK}}

	The Fokker-Planck equation corresponding to the overdamped Langevin equation [Eq.~\ref{eq:LangevinOD} in the main text] is given by 
	\begin{equation}
	\label{eq:SFPOD}
	\frac{\partial \rho}{\partial t} = \hat{\cal L}_\text{OD} = - \partial_x J_\text{OD},
	\end{equation}
where $\rho(x,t)$ is the probability density at position $x$ at time $t$, $\hat{\cal L}_\text{OD}$ is the overdamped Fokker-Planck operator, and $J_\text{OD}(x, t) = \left[ \gamma^{-1} (f_0 + f_1) - (\gamma \beta)^{-1} \partial_x\right] \rho$ is the probability current density at position $x$ at time $t$. 
	(Other symbols have the same meaning as in the main text.)
	According to Feynman-Kac theorem, the solution of the sink equation
	\begin{equation}
	\label{eq:SSink}
	\frac{\partial g}{\partial t} = \hat{\cal L}_\text{OD} g - h g,
	\end{equation}
for an arbitrary function $h(x, t)$ and initial condition $g(x, 0) = \exp{[- \beta V(x, \lambda_0)]}/Z^\text{c}(\lambda_0(0), \beta)$ is given by the following expression
	\begin{equation}
	\label{eq:SFK}
		g(x, t) = \langle \delta(x(t) - x) e^{- \int_0^\tau \mathrm{d}t \, h(t)} \rangle,
	\end{equation}
with $h(t) = h(x(t), t)$. 
	Consider now the following expression of $h$:
	\begin{equation}
	\label{eq:ShOD}
	h = h_\text{OD} = \beta \dot{\lambda}_0 \partial_{\lambda_0} V - \frac{\beta}{\gamma} \left(f_0 f_1 + \frac{1}{\beta} \partial_x f_1\right). 
	\end{equation} 
	For this particular choice of $h$, by direct substitution, we can show that the following expression of $g$ also solves the sink equation~\ref{eq:SSink}:
	\begin{equation}
	\label{eq:SunB}
	g(x, t) = \frac{e^{- \beta V(x; \lambda_0)}}{Z_0^\text{c}(A, \beta)}, \quad Z_0^\text{c}(A, \beta) = \int \mathrm{d}x e^{- \beta V(x; A)}.
	\end{equation}
Combining Eqs.~\ref{eq:SFK} and \ref{eq:SunB}, we get,
	\begin{equation}
	\label{eq:SunBFK}
	\frac{e^{- \beta V(x; \lambda_0)}}{Z_0^\text{c}(A, \beta)} = \langle \delta(x(t) - x) e^{- \int_0^\tau \mathrm{d}t \, h_\text{OD}(t)} \rangle.
	\end{equation}
	Integrating both sides of this equation with respect to $x$ at time $t = \tau$ we get
	\begin{equation}
	\label{eq:SMainOD}
	e^{- \beta \Delta F_0^\text{c}} = \frac{Z_0^\text{c}(B, \beta)}{Z_0^\text{c}(A, \beta)} = \langle e^{- \int_0^\tau \mathrm{d}t \, h_\text{OD}(t)} \rangle,
	\end{equation}
with $\Delta F_0^\text{c} = F_0^\text{c}(B, \beta) - F_0^\text{c}(B, \beta)$. 
	Equation~\ref{eq:SMainOD} is the same as Eq.~\ref{eq:ODFK} of the main text.

	\end{document}